\documentclass[twocolumn]{aastex631}
\usepackage{amsmath}
\usepackage{longtable}
\begin{document}

\title{Revisiting the Perseus Cluster II: Metallicity-Dependence of Massive Stars and Chemical Enrichment History}

\shortauthors{Leung, Walther, Yerdon, Nomoto and Simionescu}
\shorttitle{Metallicity Dependence of Si-group Elements}

\author[0000-0002-4972-3803]{Shing-Chi Leung}

\affiliation{Department of Physics, SUNY Polytechnic Institute, 100 Seymour Road, Utica, NY 13502, USA}

\author[0009-0006-8467-2163]{Seth Walther}

\affiliation{Department of Electrical and Computer Engineering, SUNY Polytechnic Institute, 100 Seymour Road, Utica, NY 13502, USA}
\affiliation{Department of Mathematics, SUNY Polytechnic Institute, 100 Seymour Road, Utica, NY 13502, USA}
\affiliation{Department of Physics, SUNY Polytechnic Institute, 100 Seymour Road, Utica, NY 13502, USA}

\author[0009-0000-5656-9659]{Henry Yerdon}

\affiliation{Department of Electrical and Computer Engineering, SUNY Polytechnic Institute, 100 Seymour Road, Utica, NY 13502, USA}
\affiliation{Department of Computer Science, SUNY Polytechnic Institute, 100 Seymour Road, Utica, NY 13502, USA}
\affiliation{Department of Physics, SUNY Polytechnic Institute, 100 Seymour Road, Utica, NY 13502, USA}

\author[0000-0001-9553-0685]{Ken'ichi Nomoto}

\affiliation{Kavli Institute for the Physics and 
Mathematics of the Universe (WPI), The University 
of Tokyo Institutes for Advanced Study, The 
University of Tokyo, Kashiwa, Chiba 277-8583, Japan}

\author[0000-0002-9714-3862]{Aurora Simionescu}

\affiliation{SRON Netherlands Institute for Space Research, Niels Bohrweg 4, 2333 CA Leiden, The Netherlands}

\affiliation{Kavli Institute for the Physics and 
Mathematics of the Universe (WPI), The University 
of Tokyo Institutes for Advanced Study, The 
University of Tokyo, Kashiwa, Chiba 277-8583, Japan}

\affiliation{Leiden Observatory, Leiden University, PO Box 9513, 2300 RA Leiden, The Netherlands}

\correspondingauthor{Shing-Chi Leung}
\email{leungs@sunypoly.edu}

\newcommand{\red}[1]{\textcolor{red}{#1}}
\newcommand{\blue}[1]{\textcolor{blue}{#1}}

\date{\today}

\submitjournal{ApJ}
\received{Jul 18 2025}
\revised{Feb 6 2026}
\accepted{Feb 21 2026}
\published{Apr 7 2026}

\begin{abstract}

The legacy Hitomi telescope has delivered the precise measurements of the chemical abundances in the Perseus Cluster, covering the Si-group (Si, S, Ar, Ca) and Fe-group elements (Cr, Mn, Ni). In Paper I (Leung et
al., ApJ 2025), we examined the role of convection parameters and presented new core-collapse supernova (CCSN) explosion models at solar metallicity, which fit the observed abundance pattern. In this article, we extend our calculation for the stellar evolutionary models and CCSN models of the initial mass $15 - 60M_{\odot}$ and the metallicity $Z = 0 - Z_{\odot}$. The detailed pre- and post-explosion chemical profiles are calculated with a large post-processing network to capture the production of $\alpha$-chain elements (e.g., Si, S, Ar), odd-number elements (e.g., P, K, Cl), and iron-group elements (e.g., Mn, Ni). We study the role of CCSNe in the production of these elements. We compare the galactic chemical evolution model based on the nucleosynthesis yield of the new massive stars and other yield tables from the literature. For each supernova yield, we perform parameter surveys and search for configurations that produce the best-fit model and best-rate model using the Perseus Cluster as the reference. From the survey, we study how individual chemical elements affect the contributions of massive stars and Type Ia supernovae in the cosmic chemical enrichment history.  

\end{abstract}

\pacs{
26.30.-k,    
}

\keywords{Supernovae (1668) -- Galaxy clusters (584) -- Perseus Cluster (1214) -- Hydrodynamical simulations (767) -- Explosive nucleosynthesis (503) -- Chemical abundances (224)}



\section{Introduction}

\subsection{Inspiration from Precise Observations}

The Hitomi Telescope (Astro-H) \citep{Takahashi2016Hitomi} is the pioneer of precise spectroscopic measurements at X-ray energies above 2~keV. The Perseus Cluster is one of the very few objects observed by this legacy telescope, which provided novel measurements of, e.g., its gas mean density, velocity distribution, resonance line, and temperature \citep{Hitomi2018PerseusGas, Hitomi2018PerseusResonance, Hitomi2018PerseusTemp}. Moreover, these observations allowed us to show that the hot gas filling this gigantic structure has effectively the same composition as the proto-solar nebula \citep{Hitomi2017PerseusSolar}. Because of the large size of galaxy clusters, the chemical abundance pattern of their intracluster gas probes a representative average over billions of supernova explosions, making it a powerful test of nucleosynthesis physics. 
The high precision of the Hitomi measurements, as compared to previous probes of the chemical enrichment of galaxy clusters obtained with lower resolution spectrographs, enables the community to better test whether existing numerical massive star models are consistent with the metal abundance pattern in the Perseus Cluster \citep{Simionescu2019Perseus}. 

The next generation XRISM (X-Ray Imaging and Spectroscopy Mission), launched in September 2023, is now following in the footsteps of Hitomi and unraveling the properties of many more astrophysical targets at similarly high spectral resolution in the high-energy X-ray band. XRISM will probe the precise chemical enrichment pattern in a number of clusters, in addition to Perseus (for example, the Centaurus and Virgo Clusters observed during the performance verification phase of the mission). It will also deliver novel data for other types of astrophysical sources relevant for understanding cosmic nucleosynthesis, such as mapping metal abundances in individual supernova remnants such as Cassiopeia~A. It is therefore an opportune moment to reexamine existing models of massive stars, which can be compared to this rich upcoming set of data.

\subsection{Stellar Population and Galactic Chemical Evolution}

The stellar populations and their chemical abundance patterns in stars across the Milky Way offer a rigorous test for supernova models. Stars at different metallicity reveal the chemical trends at different cosmic ages by stellar surveys, e.g., The APOGEE \citep[Apache Point Observatory Galactic Evolution Experiment, ][]{APOGEE2017} in the Sloan Digital Sky Survey IV \citep[SDSS][]{SDSS172022}. 

The Galactic Chemical Evolution (GCE) is a summative tool for comparing collectively stellar and supernova models with the quantitative features of the stellar population \citep{Matteucci2012GCEBook, Matteucci2016GCEReivew}. The model considers how chemical elements are produced and destroyed by generations of stars and various galactic processes. The stellar population at [Fe/H] $\sim-1-0$ is important for Type Ia supernova (SN Ia) models, while early values and trends ([Fe/H] $\sim-3$) could reflect on the massive stars. For example, in \cite{Matteicci2009GCESNIa, Kobayashi2020GCE-SNIa}, the evolution of elements like O/Fe and Mn/Fe in the Milky Way Galaxy is used to constrain the SN Ia delay-time distribution and their fraction being exploded from Chandrasekhar mass (Ch-mass) vs. sub-Chandrasekhar mass (subCh-mass) WDs. 

The low metallicity modals are also important for reconciling with the chemical abundance of the Perseus Cluster. While the Perseus Cluster has a near-solar metallicity \citep{Hitomi2017PerseusSolar}, the intra-cluster gas is a mixture of all SNe that have occurred over the last few billion years. To accurately model this, the low-Z SNe models as well as solar metallicity models are necessary components for a consistent comparison.

\subsection{Motivation}

In \cite{Simionescu2019Perseus}, the precise chemical abundance pattern of the hot gas in the Perseus Cluster measured by the legacy Hitomi telescope is reported. The abundance patterns are very similar to the solar composition, even though the hot gas is the repository of billions of stars across cosmic history. They further reported that the existing massive star and Type Ia supernova models cannot explain the observed pattern: models in general overproduce Si/Fe and S/Fe, and underproduce Ar/Fe and Ca/Fe. Given the similarity of the abundances of the Perseus Cluster with the solar composition, this suggests a similar mismatch of the current massive star + SN Ia models with the solar abundances.  

Inspired by this mismatch, in \citet[Paper I][]{Leung2025Perseus1}, we investigated the role of convection parameters in stellar evolution on the structure of massive stars, and how these parameters affect the final chemical yield. We constructed stars with the initial masses of $M_{\rm ZAMS}=15-40~M_{\odot}$ at the solar metallicity, and computed their explosive nucleosynthesis using spherical models. We tune the mixing length parameter $\alpha$ and semi-convection parameter $\alpha_{\rm SC}$ and search for the best values which give a better fit to the chemical abundance pattern of the Perseus Cluster, especially Si/Fe, S/Fe, Ar/Fe, and Ca/Fe. 

In Paper I, we show that the evolution of some Fe-group elements, e.g., Mn and odd-number elements in the Si-group elements, requires a detailed chemical composition of the pre-explosion. Thus, we developed our post-processing method for our stellar evolutionary code to capture the chemical composition at the pre-explosion phase with a larger network. This lets us keep track of minor elements, e.g., K, Sc, Mn, across the stellar life. The models with different metallicity enable us to compare consistently with other massive star models and the Perseus Cluster. We use this approach to compute models with the metallicities between $0$ and $Z_{\odot}$\footnote{In this article, the metallicity $Z$ is noted by its mass fraction without logarithm, and thus $Z = 0$ corresponds to zero metallicity and $Z_{\odot} = 0.02$ is the solar metallicity. The logarithmic one [Fe/H]=$\log_{10}$(Fe/H)/(Fe/H)$_{\odot}$ is noted explicitly.}. We want to understand how the Si-group elements vary across the metallicity. 

In Section \ref{sec:method}, we describe the numerical method for constructing the massive star explosion models. We focus on the reconstruction of detailed chemical composition in the stellar evolution models. In Section \ref{sec:results}, we present our massive star models with metallicity from zero to solar metallicity and study the trend of individual elements, including both Si-group and Fe-group elements. In Section \ref{sec:GCE}, we apply these supernova yields in the GCE. We search for the set of supernova yields where the galactic evolution can be the closest to that of the Perseus Cluster, and study the evolution of individual elements, including Si, S, Ar, Ca, Mn, and Ni. We also study their implications on supernova rates. Based on these results, we compare these rates with the observed values in the literature. In Section \ref{sec:discussion}, we discuss the caveats and possible extension of our work. At last, we give our conclusion. 

\section{Metallicity-Dependence of Stellar Models}

The metallicity dependence of stellar yields is important for explaining stellar diversity in the H-R diagrams, GCE, and stellar populations. It documents how generations of stars contribute cumulatively to reach the current level of chemical abundances. 
Multiple groups have presented their own massive star models, which span different possibilities of input physics. Some stellar models evolve up to C-burning or the onset of gravitational collapse. These models are important for reconciling with the trends of various $\alpha$ elements (such as O, Na, Mg). Below, we discuss these representative models, which fully or partially cover the massive stars ranging between $10 - 60~M_{\odot}$. 

The Henyey-type stellar evolution code reported in \cite{Nomoto1988Code, Umeda2000MassiveStar} is extensively used for building supernova progenitors. Massive star models with the zero-age main-sequence mass ($M_{\rm ZAMS}$) from $11-100~M_{\odot}$, and metallicity from $10^{-4}-0.05$ are evolved up to pre-supernova explosion \citep{Nomoto2013ARAA}. These models are widely applied in GCE \citep{Kobayashi2020GCE-SNIa, Kobayashi2020GCEUranium}, fitting on individual stars \citep{Tominaga2009Jet} and galaxies \citep{Leung2024Jet}. 

The \texttt{KEPLER} code \citep{Weaver1978Kepler} is another representative code for generating stellar models. The code directly evolves the models with a large nuclear reaction network. Large stellar model grids are reported in \cite{Woosley1995Kepler, Woosley2002Grid}, which cover stars from 0 to solar metallicity with $M=11-40~M_{\odot}$. The method is adopted for constructing a dense stellar grid in \cite{Suhkbold2016Grid}, which encompasses stars of $M=9-120~M_{\odot}$ at solar metallicity, and building zero-metallicity star models \citep{Heger2002ZeroMet}.


The FRANEC code \citep[Frascati Raphson Newton Evolutionary Code, ][]{Chieffi1989FRANEC1, Pietrinferni2004} has also been used to develop stellar evolution models covering a wide range of mass and metallicity, including low-mass stars and massive stars. The most recent models include rotation \citep[see e.g., ][]{Chieffi2004CCSN,Chieffi2013,Limongi2018Grid,Lorenzo2024GridII}, and provide stellar yields for rotating and non-rotating stars including explosive nucleosynthesis. These models aim at explaining the supernova population and early metal enrichment.

The \texttt{MESA} code \citep[Modules for the Experiments in Stellar Astrophysics, ][]{Paxton2011MESA, Paxton2013MESA,Paxton2015MESA,Paxton2018MESA,Paxton2019MESA} is a general multi-purpose and multi-physics stellar evolution code designed for simulations. In the \texttt{NuGrid} stellar data set, the chemical yields for stars of $M=1-25~M_{\odot}$ and $Z=10^{-4}-0.02$ for elements from H to Bi are presented. These yields aim at providing a theoretical counterpart to that in presolar grains studies \citep{Pignatari2016NuGridI, Ritter2018NuGridII,Battino2019NuGridIII} and GCE models \citep{Ritter2018SYGMA}. 

Massive star models with on-time optical opacity are computed with the \texttt{PARSEC} code \citep[the PAdova and TRieste Stellar Evolution Code, ][]{Bressan2012PARSECcode}. The code features the direct calculations of the opacities for any given chemical composition from scratch. Model grid of stars with masses from $M=0.1-12~M_{\odot}$ stars are constructed for $Z=5\times10^{-4}-0.07$. The models are evolved up to C burning and are designed for studying isochrones in population synthesis.

Rotating star models are also studied by the \texttt{GENEVA} code \citep{Eggenberger2008GENEVA} for various metallicity \citep[see their line of works in e.g., ][]{Ekstrom2012GENEVA1,Murphy2021GENEVA5,Yusof2022GENEVA7} for $M=0.8-300~M_{\odot}$ stars with various metallicity $Z=0-0.02$. The models up to carbon-burning are applied to, e.g., the reconstruction of the optical features of young globular clusters.

Lastly, the Cambridge STARS code \citep{Eggleton1971STARS,Pols1995STARS,Stancliffe2004STARS} is a code that features solving the stellar structure equations, nuclear reactions, and diffusive mixing simultaneously. Extensive pre-supernova models are built \citep{Eldridge2004STARS_CCSN} and are used for the binary population synthesis BPASS \citep[Binary Population and Spectral Synthesis, ][]{Eldridge2017}. The code contains the mass-loss prescription of Wolf-Rayet stars at low metallicity \citep{Dray2003WRMassLoss}.

\section{Post-Processing Nucleosynthesis}
\label{sec:method}

\begin{figure}
    \centering
    \includegraphics[width=0.95\linewidth]{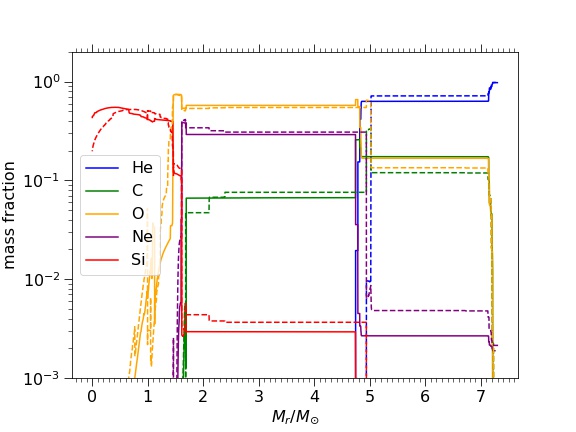}
    \caption{The chemical composition of the model with $M=20 M_{\odot}$ and $Z=Z_{\odot}$ at the end of the O burning phase by directly solving the 127-isotope network in the MESA code (solid line) and by post-processing the thermodynamics trajectory from models using the 22-isotope network (dashed line).}
    \label{fig:compare_network}
\end{figure}

In Paper I, we developed the prescription for using \texttt{MESA} (version 8118) to calculate the evolutionary paths and the chemical profiles of the massive star models, deriving the refined mixing parameters. We assume the $\alpha$-chain for the nuclear fusion network, which is useful for tracing the formation of primary isotopes. 

We note that secondary isotopes (minor isotopes depending on the initial metallicity, e.g., $^{14}$N and $^{22}$Ne) require keeping track of the seed nucleus during the evolution. A direct solution is to perform stellar evolution calculations with a large network. However, the detailed transport and synthesis of minor elements could be the bottleneck. The large network might become computationally unfeasible at a late phase, when the code attempts to resolve the detailed structure of the star. Therefore, we use the post-processing method to avoid this computational difficulty. This approximate approach is reliable when most energy is produced along the $\alpha$-chain elements, which occurs in massive stars. The energy generation and the convective motion are dictated mostly by the nuclear luminosity of these reaction channels. The production of minor elements, thus, depends on this thermodynamic background. 

Thanks to the Lagrangian formalism in \texttt{MESA}, the density and temperature of the history of the fluid shells have a one-to-one mapping to the tracers' thermodynamic history. We first define the reference grid mesh ($m_i$). At the end of each step, we map the density and temperature from the stellar model to the reference grid as a function of time. To document the convective mixing history inside the star, we record the convective parameters (the diffusion coefficient and the convective velocity) if convection exists in certain zones, which leads to mixing of chemical elements among fluid elements. 

In Figure \ref{fig:compare_network}, we show the {$20~M_{\odot}$ model with $Z=Z_{\odot}$ evolved up to the late phase of O burning. The chemical profiles are computed by two methods: (1) the model is evolved with the large 127-isotope networks fully with MESA, and (2) the model is evolved with a small 21-isotope network (identical to what we used in Paper I), but we post-process the yield using the prescription above. The two approaches agree with each other well based on the mass fraction of the elements and the zones where the elements are mixed. However, semi-convection processes, including thermohaline mixing, are not well captured near the mixing boundary.

Similar to Paper I, when the star reaches the core-collapse point at $\rho_c>10^{10}$gcm$^{-3}$, we transfer the stellar model to the 1D Lagrangian spherical explosion code using the $\alpha$-chain network. We remove the iron core, indicated by the MESA code, and inject a thermal bomb such that the star has a final energy of $10^{51}$ erg. The explosion code follows the explosion shock heating until most exothermic reactions are ceased. We proceed with the second post-processing by the thermodynamics history obtained in the explosion phase to compute the final composition. In this work, we use the same MESA inlist for stellar models of various metallicity.

In this article, we repeat the calculations by setting different initial masses ($M_{\rm ZAMS} = 15 - 60 M_{\odot}$) and initial metallicity ($Z = 0 - 0.02$). We refer the interested readers to \cite{Travaglio2004Tracer, Seitenzahl2010Tracer} for the passive tracer prescription, and \cite{Leung2015WENO} for our implementation of this prescription in our numerical code. 

\section{Results}
\label{sec:results}

\subsection{Models}

\begin{table*}[]
    \centering
    \caption{The stellar models computed in this work. $M$ and $M_f$ are the initial Zero-Age Main-Sequence (ZAMS) mass and the pre-explosion mass in units of $M_{\odot}$. $Z$ is the initial metallicity of the star. $M_{\rm CO}$, $M_{\rm Si}$, $M_{\rm Fe}$ are the C+O, Si and the Fe core mass coordinates. [P/Fe], [K/Fe], [Sc/Fe], [Mn/Fe], [Ni/Fe] are the log-10 scaled mass fractions of the element over Fe, in unit of solar value, i.e. [X/Fe] = $\log_{10}$ [(X/Fe)/(X/Fe)$_{\odot}$]. All models are noted as MAAZB, where AA is $M/M_{\odot}$ and B is $Z/Z_{\odot}$.}
    
    \begin{tabular}{c c c c c c c c c c c c c c c c}
        \hline
         Model & $M$ & $Z/Z_{\odot}$ & $M_f$ & $M^{\rm core}_{\rm CO}$ & $M^{\rm core}_{\rm Si}$ & $M^{\rm core}_{\rm Fe}$ & $M_{\rm Ni56}$ & [Si/Fe] & [P/Fe] & [K/Fe] & [Sc/Fe] & [Mn/Fe] & [Ni/Fe] \\ \hline
         M15Z0 & 15 & 0 & 3.78 & 2.26 & 2.02 & 1.38 & 0.096 & 0.29 & 0.54 & 0.22 & 0.28 & -0.44 & 0.25 \\
         M15Z1e-2 & 15 & 0.01 & 4.15 & 2.56 & 1.72 & 1.34 & 0.088 & 0.27 & 0.15 & -0.51 & -1.27 & -0.24 & 0.24 \\
         M15Z1e-1 & 15 & 0.1 & 3.88 & 2.29 & 1.67 & 1.36 & 0.088 & 0.16 & 0.14 & -0.48 & -0.58 & -0.31 & 0.22 \\
         M15Z1 & 15 & 1 & 3.68 & 2.17 & 1.68 & 1.36 & 0.081 & 0.36 & 0.70 & 0.26 & -0.03 & -0.30 & 0.33 \\ \hline
         
         M20Z0 & 20 & 0 & 6.00 & 3.71 & 1.82 & 1.38 & 0.071 & 0.52 &  0.51 & -0.21 & -0.95 & -0.03 & 0.09 \\
         M20Z1e-2 & 20 & 0.01 & 6.30 & 5.86 & 1.86 & 1.49 & 0.057 & 0.62 & 0.87 & -0.08 & -0.86 & 0.05 & 0.03 \\
         M20Z1e-1 & 20 & 0.1 & 6.10 & 3.83 & 2.07 & 1.39 & 0.133 & 0.42 & 0.01 &  -0.09 &  0.25 & -0.24 & 0.22 \\
         M20Z1 & 20 & 1 & 5.51 & 3.86 & 2.74 & 1.41 & 0.087 & 0.88 & 0.87 & 1.04 & 0.58 & -0.09 & 0.21 \\ \hline
         
         M25Z0 & 25 & 0 &  8.43 & 8.19 & 2.60 & 1.62 & 0.295 & 0.41 & 0.29 & -0.00 & -0.15 & -0.42 & 0.14 \\
         M25Z1e-2 & 25 & 0.01 & 8.90 & 8.66 & 1.99 & 1.41 & 0.098 & 0.44 & 0.30 & 0.02 & 0.42 & -0.21 & 0.84 \\
         M25Z1e-1 & 25 & 0.1 & 8.31 & 8.07 & 2.20 & 1.55 & 0.067 & 0.60 & 0.39 & 0.13 & 0.17 & 0.09 & 0.75 \\
         M25Z1 & 25 & 1 & 7.35 & 7.22 & 1.69 & 1.40 & 0.082 & 0.10 & 0.34 & -0.01 & 0.08 & -0.14 & 0.18 \\ \hline
         
         M30Z0 & 30 & 0 &  10.88 & 10.17 & 1.97 & 1.51 & 0.267 & 0.13 & 0.10 & -0.51 & -0.19 & -0.40 & 0.18   \\
         M30Z1e-2 & 30 & 0.01 & 11.47 & 11.34 & 2.04 & 1.51 & 0.239 & -0.11 & -0.17 & -0.75 & -0.10 & -0.50 & 0.19 \\
         M30Z1e-1 & 30 & 0.1 & 10.62 & 10.55 & 2.00 & 1.44 & 0.055 & 0.62 & 0.48 & 0.13 & 0.18 & 0.015 & 0.59 \\
         M30Z1 & 30 & 1 & 9.12 & 9.07 & 1.96 & 1.45 & 0.096 & 0.43 & 0.52 & 0.28 & 0.32 & -0.097 & 0.51 \\ \hline
         M40Z0 & 30 & 0 & 16.25 & 15.89 & 2.92 & 1.72 & 0.154 & 0.46 & 0.25 & -0.048 & -0.24 & -0.12 & 0.50 \\
         M40Z1e-2 & 40 & 0.01 & 16.71 & 16.69 & 3.20 & 1.51 & 0.160 & 0.74 & 0.15 & 0.076 & 0.054 & -0.059 & 0.63 \\
         M40Z1e-1 & 40 & 0.1 & 15.48 & 15.45 & 2.48 & 1.54 & 0.476 & -0.044 & -0.16 & -0.72 & -0.38 & -0.36 & -0.013 \\
         M40Z1 & 40 & 1 & 12.15 & 11.47 & 2.13 & 1.60 & 0.434 & -0.11  & 0.097 & -0.28 & 0.047 & -0.50 & 0.32 \\ \hline
         M60Z0 & 60 & 0 &  27.45 & 26.52 & 6.53 & 1.82 & 0.725 & 0.53 & 0.21 & 0.40 & -0.34 & -0.42 & 0.43 \\
         M60Z1e-2 & 60 & 0.01 & 27.77 & 27.07 & 6.26 & 1.79 & 0.845 & 0.51 & 0.00 & -0.31 & -0.45 & -0.44 & 0.35 \\
         M60Z1e-1 & 60 & 0.1 & 25.03 & 24.22 & 5.68 & 1.76 & 0.431 & 0.75 & 0.19 & 0.27 & 0.16 & -0.26 & 0.47 \\
         M60Z1 & 60 & 1 & 17.56 & 17.56 & 3.27 & 1.82 & 0.841 & 0.07 & 0.22 & -0.35 & 0.24 & -0.52 & 0.43 & \\ \hline

    \end{tabular}
    \label{tab:models}
\end{table*}

In Table \ref{tab:models} we list all the stellar evolutionary models computed in this work. The models consist of a combination of 
$M=15-60~M_{\odot}$ and $Z=0-0.02$.  Each sequence consists of metallicity from zero to solar metallicity. For the discussion purpose, we name it as MAAZB, where AA denotes $M/M_{\odot}$ and B denotes $Z/Z_{\odot}$.

The models show a wide spectrum in the C+O, Si, and Fe core masses\footnote{In MESA, the core mass is defined to be the mass coordinate where the mass fraction of indicated elements reaches below 10\%.}, $^{56}$Ni mass production, and their final abundance patterns. A higher metallicity model produces a lower core mass, especially for high mass stars, due to the enhanced mass loss at high metallicity. However, the $Z$-dependence is non-monotonic because of the dependence of H-burning on
metallicity. Examining all models, the low (but non-zero) metallicity is where the core is the most massive. The $^{56}$Ni mass shares a similar wide spread with the ejected mass between $0.08-0.8~M_{\odot}$. A star with a higher $M$ tends to explode with a larger $^{56}$Ni mass. 

The 25 $M_{\odot}$ model displays wide ranges in $^{56}$Ni. In the top panel of Figure \ref{fig:precollapse_M25}, we plot the temperature profile of models in three different $Z$. A lower metallicity results in a more extended and hotter envelope outside the Fe-core at the same mass coordinate. The bottom panel shows the chemical structure, which shows that the metallicity also affects the distribution of primary isotopes, e.g., $^{28}$Si and $^{16}$O. The M25Z0 has an expanded Si core, which allows for more extended production of $^{56}$Ni when the final explosion occurs.

The full explosive yield table for massive star models presented in this work is available at \href{https://doi.org/10.5281/zenodo.17185967}{zenodo}.

\begin{figure} 
    \centering
    \includegraphics[width=0.98\linewidth]{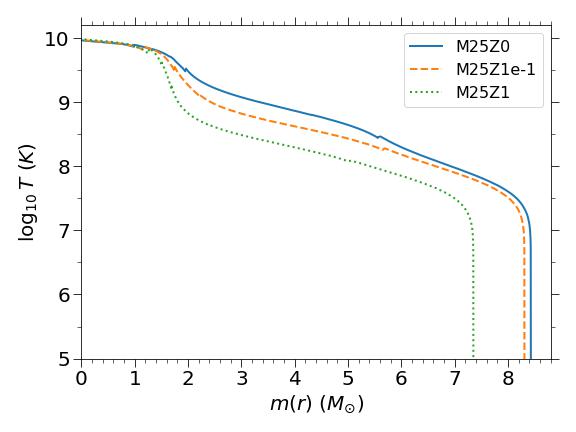}
    \includegraphics[width=0.98\linewidth]{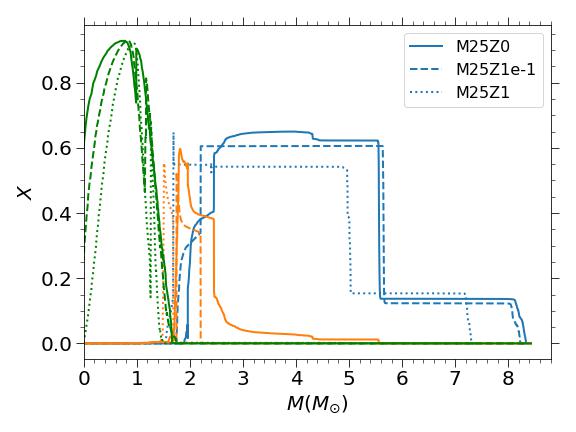}
    \caption{(top panel) The pre-collapse temperature profiles of M25Z0 (blue solid line), M25Z1e-1 (orange dashed line) and M25Z1 (green dotted line). 
    (bottom panel) Same as the top panel but for the abundances of $^{16}$O (blue), $^{28}$Si (orange) and $^{56}$Fe (green). The line style corresponds to the initial metallicity.}
    \label{fig:precollapse_M25}
\end{figure}

\subsection{Isotopic Yields}

In this section, we compare how the isotopic production depends on its initial metallicity.

In the top panel of Figure \ref{fig:xiso_M15Z} we plot the post-explosion chemical abundance [X/$^{56}$Fe]\footnote{[X/$^{56}$Fe] is the logarithmic mass fraction for the isotopes defined as [X/$^{56}$Fe] = $\log_{10}$ ((X/$^{56}$Fe)/(X/$^{56}$Fe)$_{\odot}$).} of the stable isotopes for the M15 series after all short-lived radioactive isotopes have decayed. The dependence on metallicity for this series of models is weak, except for isotopes including $^{22}$Ne, $^{31}$P, $^{45}$Sc, $^{48}$Cr, $^{50}$V. The variations of those isotopes are not monotonic due to the competition of mass loss, which reduces the core mass and hence the final compactness upon explosion, and the seed isotopes in the initial composition. Certain isotopes are overproduced, including $^{30}$Si, $^{31}$P, $^{48}$Cr and $^{64}$Ni. 

In the middle panel of the same figure, we plot the post-explosion chemical abundance of the stable isotopes for the M20 series. The sequence shows monotonic variation when $Z$ increases for most elements. A high $Z$ results in stronger production in even-number species along $\alpha$-chain from Si to Cr, and some minor elements, e.g., P, Cl, and K. The production of these elements is significantly above the solar values for the $Z=Z_{\odot}$ model. Some elements, like Ne and Mg, show an opposite trend. There is no significant change for Fe-group elements. The new models produce Sc, Mn, and Co close to the solar value.  

In the bottom panel, we plot the post-explosion chemical abundance of the stable isotopes for the M25 series. Unlike the previous two series of models, there is no clear monotonic dependence of the isotopic elements on $Z$. The $0.1~Z_{\odot}$ appears to be more prominent in the isotope production, which is super-solar. The $Z=0$ and $Z_{\odot}$ models produce isotopic ratios comparable with the solar values. Odd-number elements, including P, Cl, K, Sc, Mn, and Co, are within the two horizontal bars, suggesting that these stellar models can robustly produce these minor elements. 

\begin{figure*}
    \centering
    \includegraphics[width=0.98\linewidth]{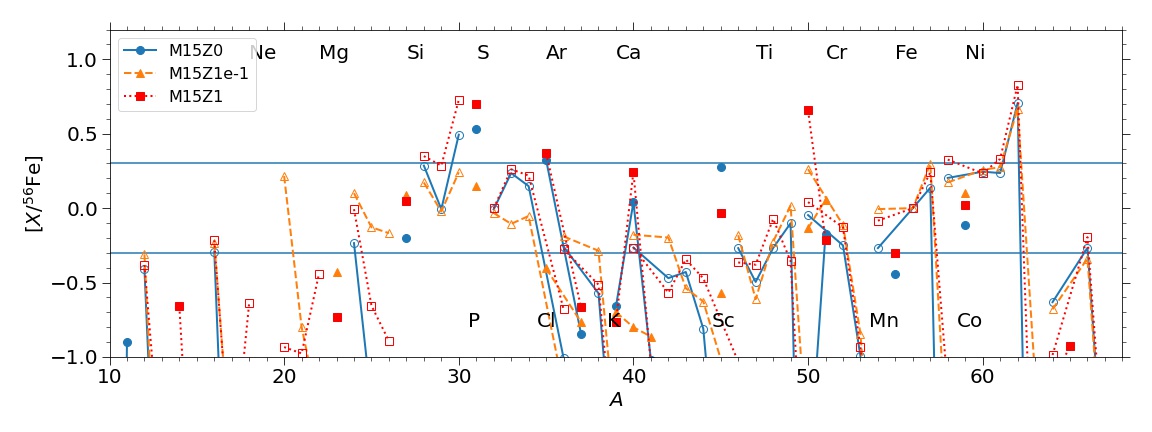}
    \includegraphics[width=0.98\linewidth]{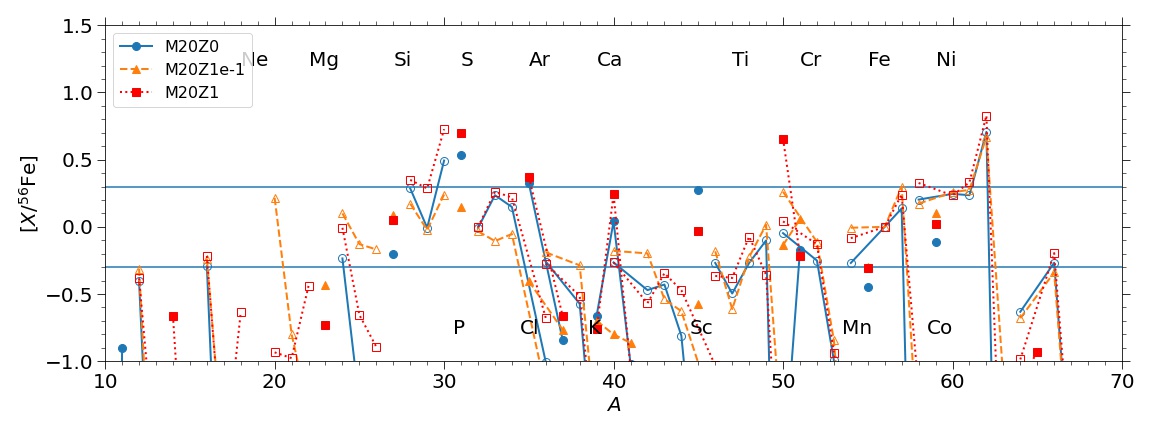}
    \includegraphics[width=0.98\linewidth]{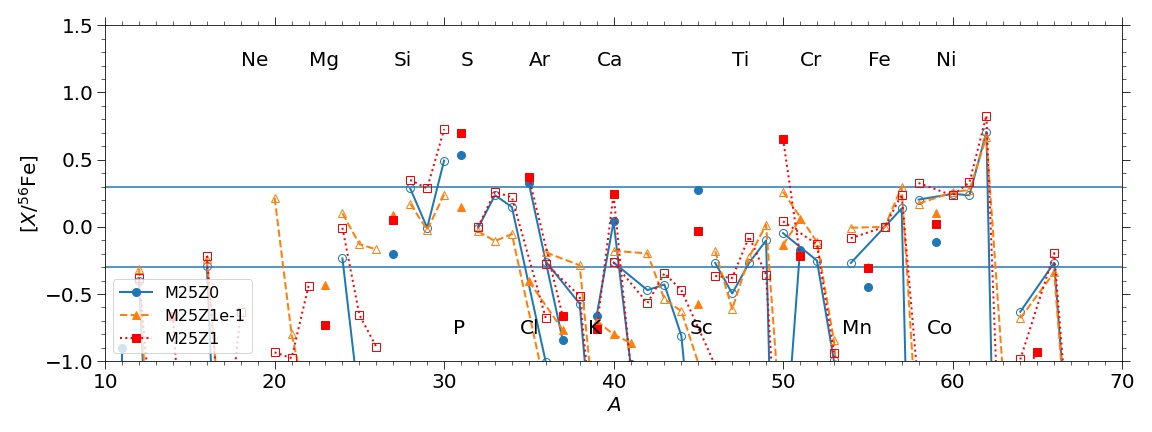}
    \caption{(top panel) The scaled mass fraction[X/$^{56}$Fe] for the stable isotopes  after explosion of 15 $M_{\odot}$ progenitor assuming $1 \times 10^{51}$ erg, with $Z = 0$ (M15Z0, blue circles), $Z = 0.1 Z_{\odot}$ (M15Z1e-1, orange triangles) and $Z = Z_{\odot}$ (M15Z1, black squares). Isotopes from C to Zn are shown. The two horizontal lines refer to two times (upper line) and half (lower line) of the solar ratios. 
    (middle panel) Same as the top panel, but for M20Z0, M20Z1e-1, and M20Z1. 
    (bottom panel) Same as the top panel, but for M25Z0, M25Z1e-1, and M25Z1.
    }
    \label{fig:xiso_M15Z}
\end{figure*}

\begin{figure*}
    \centering
    \includegraphics[width=0.98\linewidth]{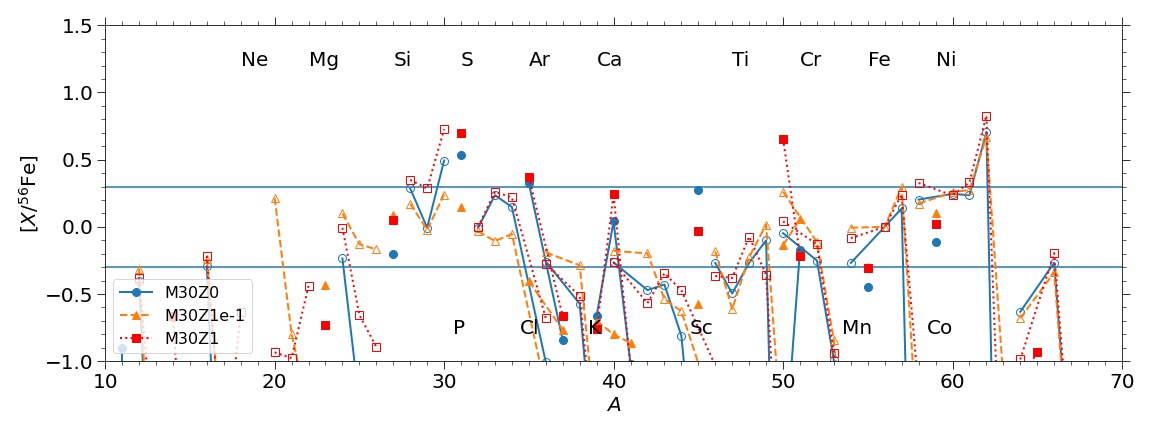}
    \includegraphics[width=0.98\linewidth]{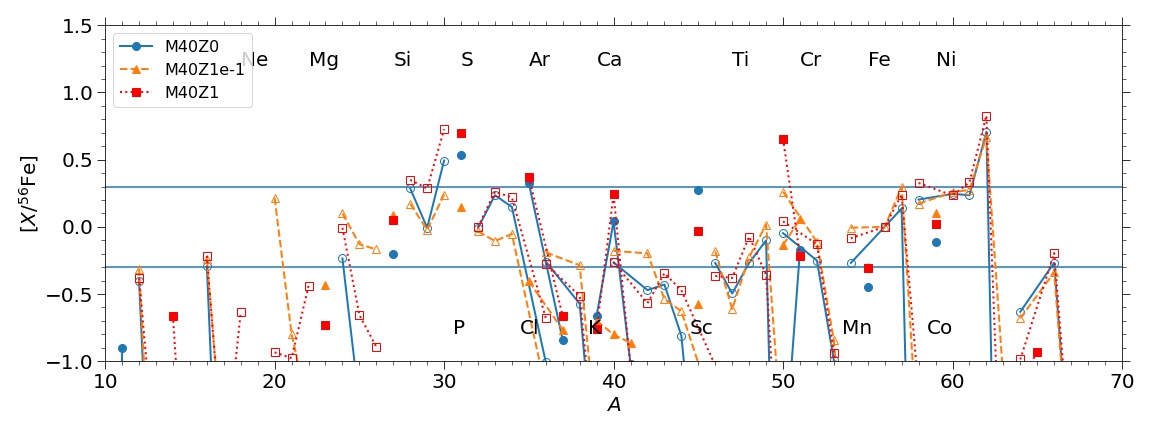}
    \includegraphics[width=0.98\linewidth]{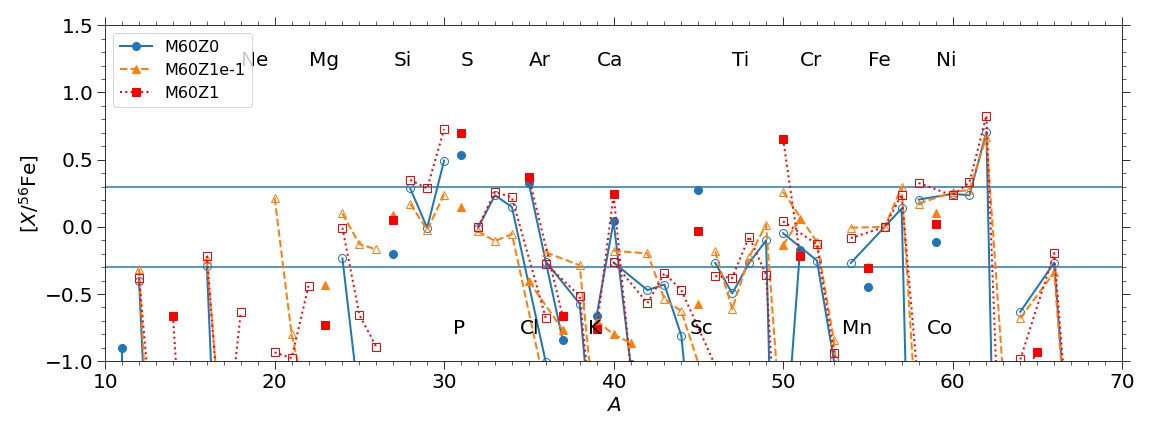}
    \caption{(top panel) The scaled mass fraction[X/$^{56}$Fe] for the stable isotopes  after explosion of 40 $M_{\odot}$ progenitor assuming $1 \times 10^{51}$ erg, with $Z = 0$ (M30Z0, blue circles), $Z = 0.1 Z_{\odot}$ (M30Z1e-1, orange triangles) and $Z = Z_{\odot}$ (M30Z1, black squares). Isotopes from C to Zn are shown.
    (middle panel) Same as the top panel, but for M40Z0, M40Z1e-1, and M40Z1. 
    (bottom panel) Same as the top panel, but for M60Z0, M60Z1e-1, and M60Z1.
    }
    \label{fig:xiso_M40Z}
\end{figure*}

In the top panel of the Figure \ref{fig:xiso_M40Z}, we plot the post-explosion chemical abundance of the stable isotopes for the M30 series. The sequence shows large variations when $Z$ increases for Si-group elements. From zero to 0.1 $Z_{\odot}$, a sharp increase of individual productions due to the metallicity effect, where the isotopic productions are significantly above their Solar values. But as $Z$ further increases, the mass loss overturns the trend, which results in a lower production. Similar trends can be seen for Fe-group elements at a lower level. The new models produce Sc, Mn, and Co close to the solar value.  

The middle panel focuses on the M40 series. The explosion favours the production of light elements and decreases with atomic mass, except for Ni and Zn. The sequence shows opposite Z-dependence compared to the M30 series. A high $Z$ results in a higher production in even-number elements at high metallicity, and an opposite trend appears at low metallicity. The production of these elements in the zero metallcity model is significantly higher than the solar values. Fe-group elements at various metallicity continue to be similar among models.

In the bottom panel, we show the chemical abundance of the M60 series. The models are similar to the M30 series, but the solar metallicity models are further suppressed due to a higher mass loss. The lower metallicity models produce super-solar ratios for most elements, while the $Z_{\odot}$ model is comparable with the solar composition. Odd-number elements, including P, Cl, K, Sc, Mn, and Co, agree with the solar values, suggesting that keeping track of the detailed abundance allows us to robustly produce these elements naturally. 

\subsection{Elemental Yields}

\begin{figure*}
    \centering
    \includegraphics[width=0.48\linewidth]{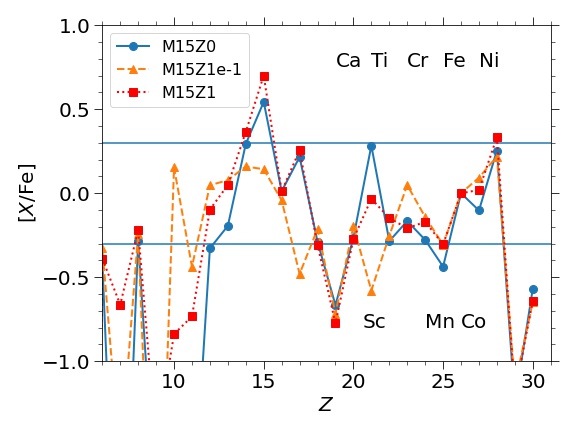}
    \includegraphics[width=0.48\linewidth]{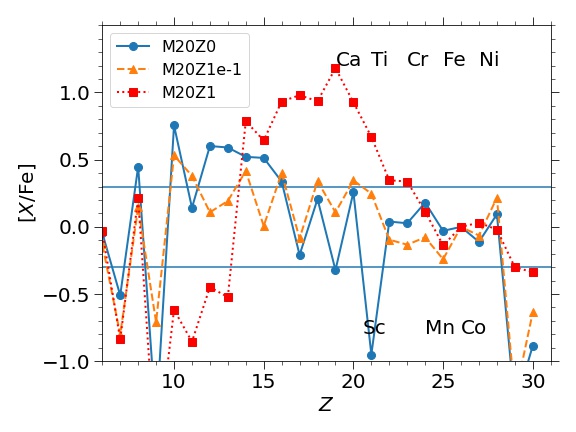}
    \includegraphics[width=0.48\linewidth]{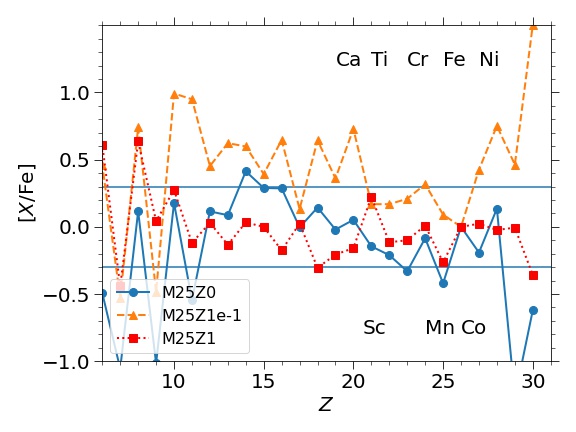}
    \includegraphics[width=0.48\linewidth]{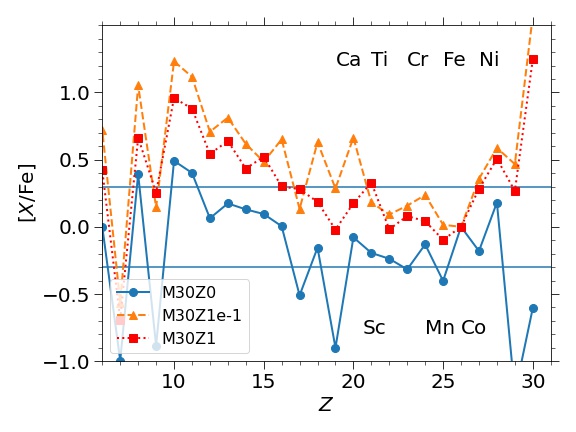}
    \includegraphics[width=0.48\linewidth]{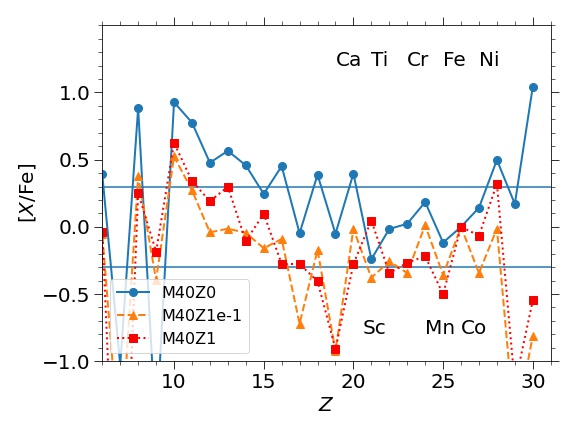}
    \includegraphics[width=0.48\linewidth]{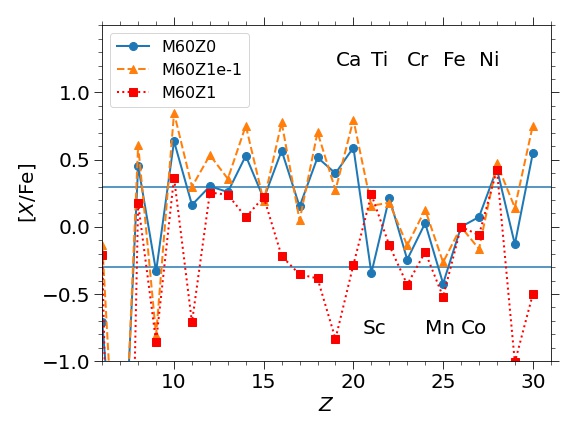}
    \caption{(top left panel) The scaled mass fraction[X/Fe] for the stable isotopes  after explosion of 15 $M_{\odot}$ progenitor assuming $1 \times 10^{51}$ erg, with $Z = 0$ (blue circles), $Z = 0.1 Z_{\odot}$ (orange triangles) and $Z = Z_{\odot}$ (black square). Elements from C to Zn are shown.
    Other panels are the same as the top left panel, 
    but for M20Z0, M20Z1e-1, M20Z1 (top right panel), 
    M25Z0, M25Z1e-1, M25Z1 (middle left panel), 
    M30Z0, M30Z1e-1, M30Z1 (middle right panel), 
    M40Z0, M40Z1e-1, M40Z1 (bottom left panel), 
    M60Z0, M60Z1e-1, M60Z1 (bottom right panel).
    }
    \label{fig:xiso_ele}
\end{figure*}

In Figure \ref{fig:xiso_ele} we plot the elemental yields of our model sequences for $M = 15 M_{\odot}$ (top left panel), $M = 20 M_{\odot}$ (top right panel), $M = 25 M_{\odot}$ (middle left panel), $M = 30 M_{\odot}$ (middle right panel), $M = 40 M_{\odot}$ (bottom left panel), and $M = 60 M_{\odot}$ (bottom right panel). 

The 15 $M_{\odot}$ models show most elements agreeing with the Solar values at different metallicity, except for the overproduction of P. Most elements are not sensitive to the metallicity. This is because the overall mass loss in this mass range is smaller than that of higher mass stars. While metallicity changes the final mass and the core masses, the C+O core among the computed models remains similar to each other.
The odd-even element differences are big, which can exceed 0.5 dex. 

For the 20, 25, 30, and 40 $M_{\odot}$ series, a consistent trend shows that lighter elements are more abundant than the heavy elements. Most $\alpha$-chain elements are above solar values at low metallicity, but are solar or sub-solar at $Z_{\odot}$. The odd-even element differences are seen, but with smaller intervals. Some models, such as M20Z1, show distinctive abundances that reflect the occasional shell merger events in specific masses. In non-zero metallicity models and M40Z0, [Zn/Fe] has a high value between 0.5 and 1.5. The high Zn production at low metallicity could be important for some Zn-enriched metal-poor stars. These stars could be likely enriched by single or a few supernova explosions \citep{Hartwig2023AIFirstStar}, or the aspherical explosions of massive stars \citep[see e.g., ][]{Woosley2006Collapsar,Tominaga2007,Tominaga2009Jet,Leung2023Jet1}.

At last, in the 60 $M_{\odot}$ series, the low metallicity models show a persistent overproduction for elements from Ne to Ca. Fe-group elements agree with the solar values. A small overproduction of Zn is also observed.  

We remarked that the shell merger is a rare phenomenon in massive star models. For stars undergoing core C- or O-burning, the overshooting region of the C/O convective zone can connect with the convective zone in the He-shell. This triggers an extensive mixing region where fresh $^{4}$He is brought to the hot C/O-shells. This amplifies substantially the nuclear luminosity and minor odd-Z elements by $p$-capture. \cite{Roberti2025ShellMerger} examined massive star models from different stellar evolution codes, and they observed that shell mergers persistently occur in a small fraction of models regardless of rotation. The shell merger in general occurs in stars with a small CO core ($2-5~M_{\odot}$) and low $X^{12}$C fraction (up to 0.28), where the low $^{12}$C fraction reduces the entropy discontinuity from the C-shell, and hence the mixing region could extend beyond those shells. The shell merger also appears in other \texttt{MESA} models \citep[e.g., ][]{Ritter2018NuGridII, Wu2021Wave, Leung2021Wave}.

\section{Applications: Galactic Chemical Evolution}
\label{sec:GCE}

\subsection{Model Description and Parameter Survey}

We use the Galactic Chemical Evolution (GCE) code documented in \cite{Timmes1995GCE}. The code solves the one-zone model, which integrates the effects of low-mass stars (by novae, SNe Ia, stellar wind) and high-mass stars (by supernova explosion) on the production and consumption of chemical elements over cosmic history. In this work, we substitute the default CCSN yields \citep{Woosley1995Kepler} and SN Ia yields \citep{Nomoto1984W7,Woosley1994SubChand} to study how the new models presented in this work change the evolution of individual elements. We are interested in the Si-group elements (Si, S, Ar, Ca) and some Fe-group elements (Mn and Ni), which are precisely measured in the Perseus Cluster. We remind that the Fe-group elements, especially Mn, are mostly produced in SNe Ia \citep{Seitenzahl2013SNIaYields, Nomoto2018HBSN, Leung2023SNIaReview}, especially from Ch-mass WDs.

In Table \ref{tab:bestfit} we list the models used in our GCE calculation. Each model is categorized by two types of input, including the supernova input and the model parameters. The supernova models include the chemical yields from this work and from the literature, including those for the massive stars, the Ch-mass WD models for SNe Ia, and the subCh-mass WD models for SNe Ia as described below. We use two model parameters as independent variables: the fraction of SN Ia in the stellar population $f_{\rm Ia}$, and the fraction of Ch-mass WD $f_{\rm Chand}$. These parameters conform with the parameter study in \cite{Simionescu2019Perseus}. In the default model reported in \cite{Timmes1995GCE}, the value $f_{\rm Ia} = 0.007$ can match the observed SN fraction between Type Ia and Type II. We define the `best-fit' model having the minimum $\chi^2$ from the Perseus cluster measurement for the element ratios [Si/Fe], [S/Fe], [Ar/Fe], [Ca/Fe], [Mn/Fe] and [Ni/Fe], and `best-rate' model having the closest SN Ia rate and Ch-mass WD fraction compared to transient surveys (see Section \ref{sec:rates}).

The models are named by the massive star models and the SN Ia models. We choose representative models with different explosion mechanisms and progenitors to maximize our span in the theoretical models. Each SN Ia is chosen with $^{56}$Ni $\sim 0.6~M_{\odot}$ to represent the average SNe Ia observed. For massive stars, we have included NKT13 \citep{Nomoto2013ARAA}, CL04 \citep{Chieffi2004CCSN} and WW95 \citep{Woosley1995Kepler}. For Ch-mass SN Ia yields, we include LN18 \citep[Model 300-1-c3-1][]{Leung2018ChandIa}, TM16 \citep[T1.4, ][]{Townsley2016SNIa, Keegans2023SNIaYields}, W7 \citep{Nomoto1984W7}, LC22 \citep{Lach2022GCD}, and SC13 \citep[Model N100, ][]{Seitenzahl2013SNIaYields}. For subCh-mass SN Ia yields, we include SK18 \citep[S1.0, ][]{Shen2018DDDDDD}, PK13 \citep[1.1+0.9, ][]{Pakmor2013Merger}, WW94 \citep[07, ][]{Woosley1994SubChand} and GC21 \citep{Gronow2021SubChand}. The default yield in the original GCE code uses WW95, W7, and WW94.

For LN18, we computed an extra model, LN18-Ka4, which assumes the same deflagration-detonation transition model, but the detonation transition is set at Kalorvitz number $ = 4$\footnote{The Kalorvitz is a dimensionless parameter $\textrm{Ka} = l_{\rm turb}/l_{flame}$ which controls the deflagration-detonation transition condition. See e.g., \cite{Niemeyer1995TurbFlame}.}. The resultant model has a similar Mn/Fe ratio but a lower Ni/Fe ratio. We use this extra model to contrast the weight of Ni production to the best-fit parameter.

\begin{table*}[]
    \centering
    \caption{The GCE models computed in this work and the corresponding best-fit and default parameters. The columns ``Massive Stars', `Ch-mass', and `subCh-mass' correspond to the supernova models for massive stars, Chandrasekhar mass WDs, and sub-Chandrasekhar mass WDs. For the massive stars, they include this work, NKT13 \citep{Nomoto2013ARAA}, CL04 \citep{Chieffi2004CCSN}, and WW95 \citep{Woosley1995Kepler}. For Chandrasekhar Type Ia yields, we include LN18 \citep[Model 300-1-c3-1][]{Leung2018ChandIa} and its variation with the same model but delayed detonation (Kalorwitz number = 4, LN18-Ka4), TM16 \citep[T1.4, ][]{Townsley2016SNIa, Keegans2023SNIaYields}, W7 \citep{Nomoto1984W7}, LC22 \citep{Lach2022GCD}, and SC13 \citep[Model N100, ][]{Seitenzahl2013SNIaYields}. For Sub-Chandrasekhar Type Ia yields, we include SK18 \citep[S1.0, ][]{Shen2018DDDDDD}, PK13 \citep[1.1+0.9, ][]{Pakmor2013Merger}, WW94 \citep[07, ][]{Woosley1994SubChand} and GC21 \citep{Gronow2021SubChand}. The GCE parameters $f_{\rm Ia}$ and $f_{\rm Chand}$ correspond to the fraction of SNe Ia and the fraction of Chandrasekhar mass WD within the SNe Ia class. $\chi^2$ measures the deviation from the abundance of the Perseus Cluster. [X/Fe] is the scaled mass fraction at the current universe predicted by that model.}
    \begin{tabular}{c c c c c c c c c c c }
        \hline
         Model & Massive Stars & Ch-mass & subCh-mass & $f_{\rm Ia}$ & $f_{\rm Chand}$ & $\chi^2$ (min)  & [Si/Fe] & [S/Fe] & [Ar/Fe] & [Ca/Fe] \\ \hline
        L25-LN18 & this work & LN18 & WW94 & 0.0245 & 0.23 & 27.68 & -0.020 & -0.078 & -0.109 & -0.109\\
        L25-LN18(Ka4) & this work & LN18 & WW94 & 0.0305 & 0.28 & 21.67 & -0.027 & -0.081 & -0.116 & -0.099 \\
        L25-SC13 & this work & SC13 & WW94 & 0.0250 & 0.20 & 24.77 & -0.007 & -0.079 & -0.114 & -0.111 \\
        L25-TM16 & this work & TM16 & WW94 & 0.0205 & 0.30 & 29.52 & -0.021 & -0.082 & -0.109 & -0.104\\
        L25-LC22 & this work & LC22 & WW94 & 0.0370 & 0.64 & 14.46 & 0.074 & -0.043 & -0.104 & -0.089\\
        L25-W7 & this work & W7 & WW94 & 0.0185 & 0.85 & 16.86 & -0.008 & -0.718 & -0.095 & -0.131\\
        L25-SK18 & this work & W7 & SK18 & 0.0240 & 0.42 & 7.91 & 0.010 & -0.053 & -0.083 & -0.094 \\
        L25-PK13 & this work & W7 & PK13 & 0.0175 & 1.00 & 17.11 & -0.003 & -0.068 & -0.091 & -0.135\\
        L25-GC21 & this work & W7 & GC21 & 0.0435 & 0.00 & 9.705 & -0.041 & -0.082 & -0.126 & -0.050 \\ \hline
        NKT13-LN18 & NKT13 & LN18 & WW94 & 0.0105 & 0.27 & 18.22 & 0.016 & -0.071 & -0.122 & -0.067 \\
        NKT13-LN18(Ka4) & NKT13 & LN18 & WW94 & 0.0125 & 0.30 & 14.65 & 0.011 & -0.071 & -0.122 & -0.059 \\
        NKT13-SC13 & NKT13 & SC13 & WW94 & 0.0105 & 0.23 & 16.41 & 0.032 & -0.069 & -0.123 & -0.066\\
        NKT13-TM16 & NKT13 & TM16 & WW94 & 0.0085 & 0.39 & 19.21 & 0.019 & -0.074 & -0.123 & -0.061\\
        NKT13-LC22 & NKT13 & LC22 & WW94 & 0.0145 & 0.59 & 15.42 & 0.080 & -0.046 & -0.115 & -0.055\\
        NKT13-W7 & NKT13 & W7 & WW94 & 0.0080 & 1.00 & 9.588 & 0.024 & -0.070 & -0.114 & -0.087\\
        NKT13-SK18 & NKT13 & W7 & SK18 & 0.0090 & 0.71 & 8.283 & 0.032 & -0.061 & -0.107 & -0.071\\
        NKT13-PK13 & NKT13 & W7 & PK13 & 0.0080 & 1.00 & 9.587 & 0.025 & -0.070 & -0.114 & -0.087\\
        NKT13-GC21 & NKT13 & W7 & GC21 & 0.0140 & 0.21 & 7.982 & -0.001 & -0.075 & -0.127 & -0.043\\ \hline
        CL04-LN18 & CL04 & LN18 & WW94 & 0.0135 & 0.15 & 16.00 & 0.006 & -0.071 & -0.110 & -0.056\\
        CL04-LN18(Ka4) & CL04 & LN18 & WW94 & 0.0160 & 0.20 & 13.94 & -0.005 & -0.077 & -0.118 & -0.056\\
        CL04-SC13 & CL04 & SC13 & WW94 & 0.0140 & 0.14 & 14.82 & 0.010 & -0.075 & -0.117 & -0.060\\
        CL04-TM16 & CL04 & TM16 & WW94 & 0.0120 & 0.19 & 16.46 & 0.008 & -0.073 & -0.110 & -0.051\\
        CL04-LC22 & CL04 & LC22 & WW94 & 0.0195 & 0.52 & 10.92 & 0.059 & -0.055 & -0.114 & -0.056\\
        CL04-W7 & CL04 & W7 & WW94 & 0.0110 & 0.79 & 9.218 & 0.009 & -0.074 & -0.108 & -0.080\\
        CL04-SK18 & CL04 & W7 & SK18 & 0.0135 & 0.51 & 5.294 & 0.018 & -0.062 & -0.100 & -0.062\\
        CL04-PK13 & CL04 & W7 & PK13 & 0.0105 & 1.00 & 9.489 & 0.010 & -0.074 & -0.108 & -0.087\\
        CL04-GC21 & CL04 & W7 & GC21 & 0.0195 & 0.19 & 7.626 & -0.017 & -0.079 & -0.124 & -0.041\\ \hline
        WW95-LN18 & WW95 & LN18 & WW94 & 0.0165 & 0.19 & 20.30 & 0.010 & -0.074 & -0.105 & -0.086\\
        WW95-LN18(Ka4) & WW95 & LN18 & WW94 & 0.0200 & 0.24 & 17.07 & -0.001 & -0.079 & -0.1131 & -0.0821\\
        WW95-SC13 & WW95 & SC13 & WW94 & 0.0170 & 0.17 & 18.69 & 0.016 & -0.077 & -0.111 & -0.089\\
        WW95-TM16 & WW95 & TM16 & WW94 & 0.0145 & 0.24 & 21.00 & 0.009 & -0.078 & -0.106 & -0.083\\
        WW95-LC22 & WW95 & LC22 & WW94 & 0.0250 & 0.60 & 13.23 & 0.072 & -0.053 & -0.110 & -0.077\\
        WW95-W7 & WW95 & W7 & WW94 & 0.0130 & 0.85 & 12.76 & 0.016 & -0.074 & -0.099 & -0.110\\
        WW95-SK18 & WW95 & W7 & SK18 & 0.0165 & 0.47 & 7.973 & 0.026 & -0.061 & -0.091 & -0.084\\
        WW95-PK13 & WW95 & W7 & PK13 & 0.0125 & 1.00 & 12.90 & 0.018 & -0.073 & -0.098 & -0.115\\
        WW95-GC21 & WW95 & W7 & GC21 & 0.0280 & 0.03 & 8.876 & -0.017 & -0.081 & -0.1231 & -0.046\\ \hline

    \end{tabular}
    \label{tab:bestfit}
\end{table*}

\begin{figure*}
    \centering
    \includegraphics[width=0.48 \textwidth]{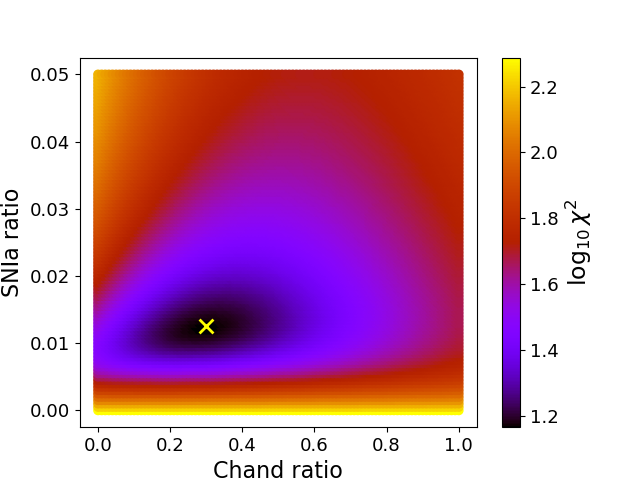}
    \includegraphics[width=0.48 \textwidth]{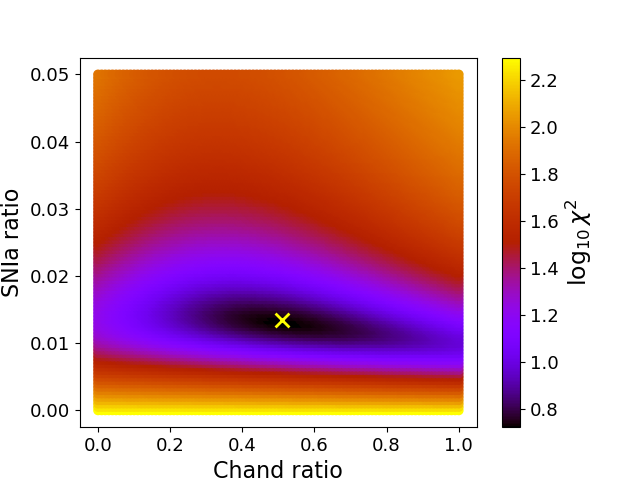}
    \includegraphics[width=0.48 \textwidth]{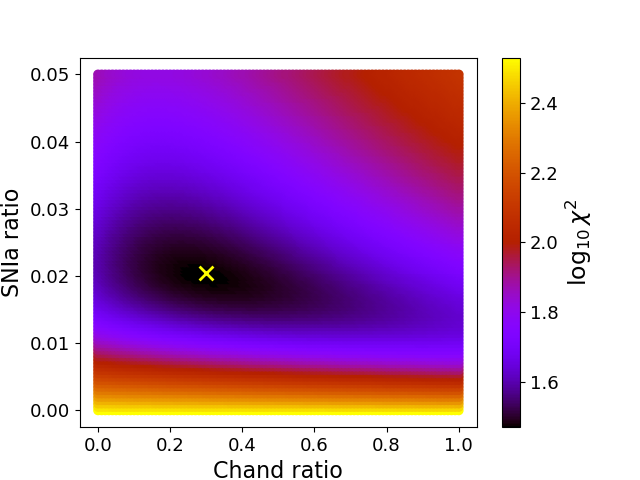}
    \includegraphics[width=0.48 \textwidth]{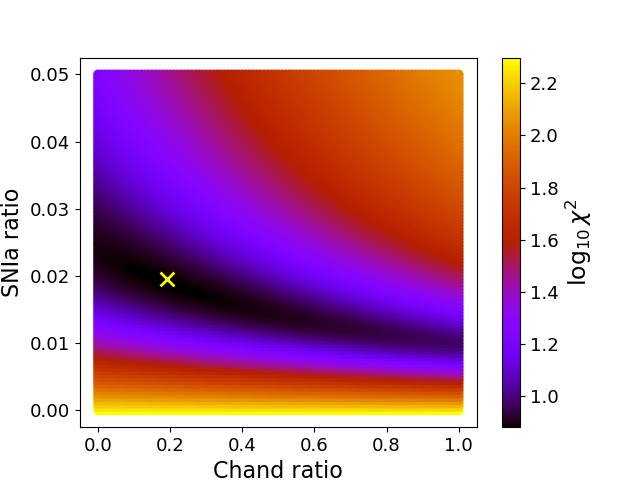}
    \includegraphics[width=0.48 \textwidth]{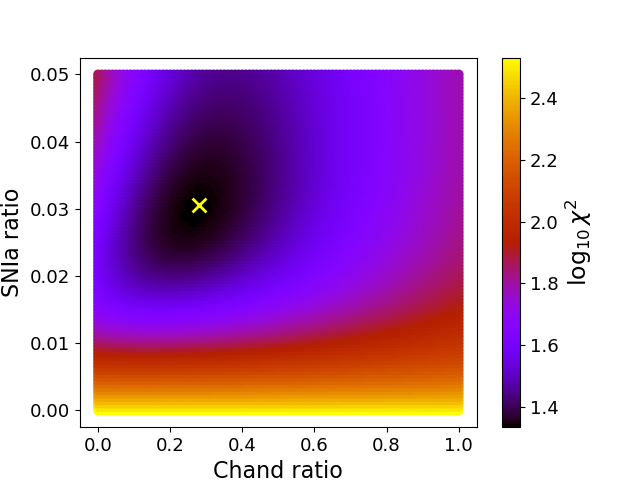}
    \includegraphics[width=0.48 \textwidth]{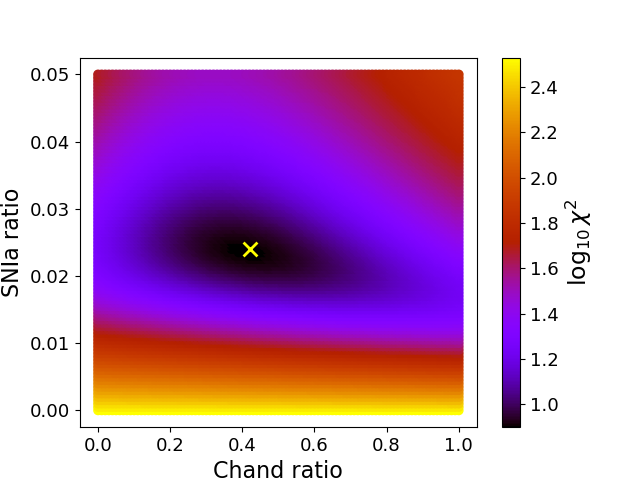}
    \caption{The left column shows the best-rate models, represented by the three models with SNe~Ia fractions $f_{\rm Ia}$ and $f_{\rm Chand}$ closest to the empirical values of 0.007 and 0.33, respectively. The right column shows the lowest $\chi^2$ best-fit models. In both columns, models are arranged from best to worst from top to bottom. The yellow cross corresponds to the parameters for the best-fit model. The left column contains the $\chi^2$-fitting of the best-rate models, including NKT13-LN18(Ka4) (top left panel),   CL04-SK18. (middle left panel) and L25-LN18(Ka4) (bottom left panel). The right columns contain those of the best-fit models, including CL04-SK18 (top right panel),  CL04-GC21 (middle left panel), and L25-SK18 (bottom right panel).}
    
    \label{fig:GCE_chisq}
\end{figure*}

In Figure \ref{fig:GCE_chisq} we show the $\chi^2$ color-plot for our GCE models with the minimum $\chi^2$ or the closest pair ($f_{\rm Ia}$ and $f_{\rm Chand}$). NKT13-LN18(Ka4),  L25-TM16, L25-LN18 have the smallest $\chi^2$ while CL04-SK18, CL04-GC21 and L25-SK18 have the closest SN rates. 

In each plot, we fixed the same massive star and Type Ia models. We vary the $f_{\rm Ia}$ and $f_{\rm Chand}$ and run the GCE model and search for the best-fit parameter with the closest composition as the Perseus Cluster, including Si, S, Ar, Ca, Mn, and Ni. The contour shape of the $\chi^2$ depends strongly on the supernova yields. Most models give a good fit of $\chi^2$ about $7-30$, meaning that a total of $0.4-0.9 \sigma$ per degree of freedom. Most models, such as NKT13-LN18, show a localized best fit, while some, e.g., CL04-GC21, show a wide band for the best-fit model. 

The implied $f_{\rm Chand}$ depends strongly on the SN Ia models but weakly on CCSN models. For example, the $f_{\rm Chand}$ rate varies between $0.15-0.27$ for models using the LN18 SN Ia yield with different massive star models. Meanwhile, the rate could vary between 0 and 1 for the same L25 massive star models but different SN Ia models. The implied $f_{\rm Ia}$, on the other hand, is less sensitive and ranges about $0.02-0.03$ of the stellar population. That is, about $2-3\%$ of the stars from $3-8~M_{\odot}$ in binary systems explode as SNe Ia. In \cite{Timmes1995GCE}, this rate is reported to be 0.7\% to match the SN statistics.

The actual fitting is very close to the Solar values, and hence to the Perseus Cluster. Shown in Table \ref{tab:bestfit}, the chemical abundances of the best-fit models and the least-fit models only differ by at most 0.1 dex. Meanwhile, the $\chi^2$ fitting depends on the selected SN Ia model. This suggests that future fitting of the Perseus Cluster depends on improved SN Ia models.

\subsection{Comparison with the Perseus Cluster}

\begin{figure*}
    \centering
    \includegraphics[width=0.48\linewidth]{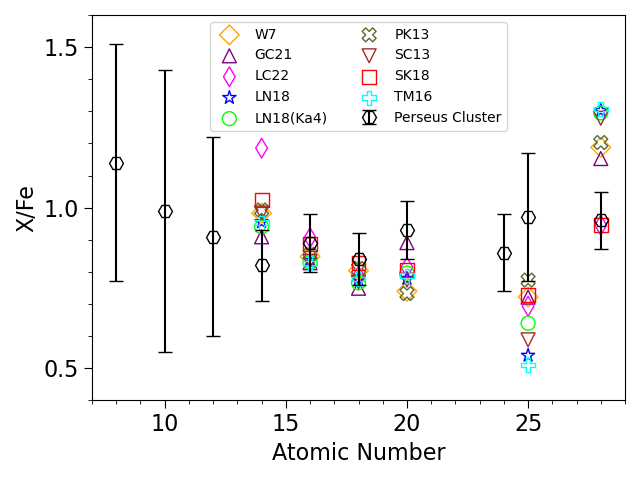}
    \includegraphics[width=0.48\linewidth]{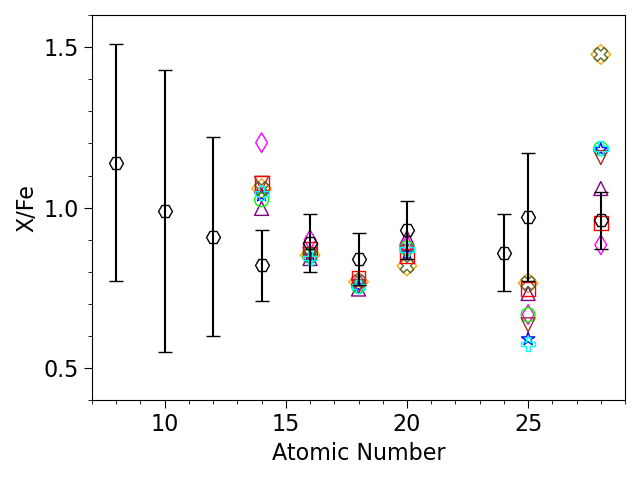}
    \includegraphics[width=0.48\linewidth]{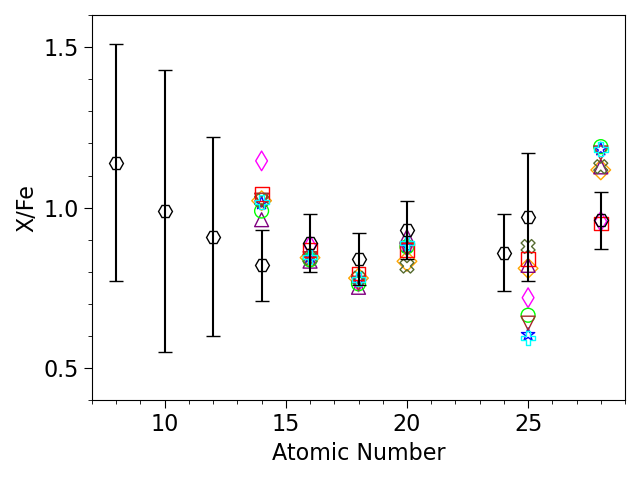}
    \includegraphics[width=0.48\linewidth]{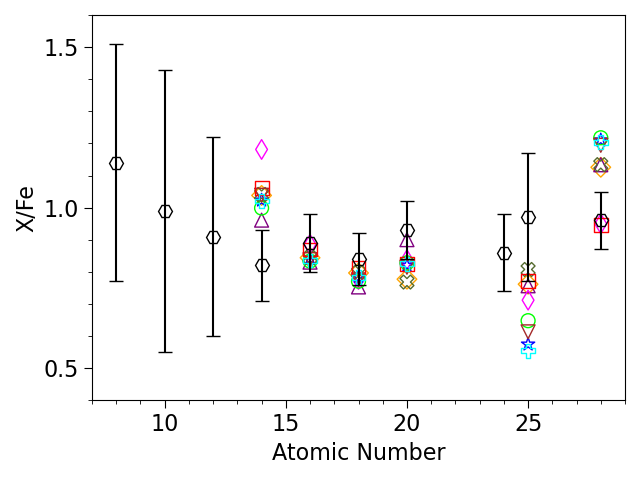}
    \caption{
    The chemical abundance X/Fe for the best models with L25-series (top left panel), with different SN Ia models including  LN18 (blue stars), LN18(Ka4) (lime circles), SC13-bestfit (brown triangles), TM16-bestfit (cyan pluses), LC22-bestfit (magenta diamonds), W7-bestfit (orange diamonds), SK18-bestfit (red squares), PK13-bestfit (green X), GC21-bestfit (purple triangles). The error bars correspond to the measurements of the Perseus Cluster from \cite{Simionescu2019Perseus}.
    The top right panel, bottom left, and bottom right panels are the same as the top left panel, but for NK13, CL04, and WW95, respectively.
    }
    \label{fig:yield_bestfit}
\end{figure*}

To further quantify the fitting of the Perseus Cluster abundances, we plot the final abundances of our GCE models and compare them with those of the Perseus Cluster. 

In the top left panel of Figure \ref{fig:yield_bestfit}, we compare the best-fit yields by fixing the massive star models to be L25 and varying the SN Ia models.  The overproduction of Si is reduced except for LC22. Both S and Ar are closer to the observed values. Ca is improved but remains about $1\sigma$ below the actual value. All models predict Mn/Fe within 1 $\sigma$ (Standard Deviation) below the lower limit, which is the result of compensating for the high Ni yield in the Ch-mass SN Ia model. As we noted, due to the precise measurement in Ni/Fe (the uncertainty of Ni/Fe is one-third of Mn/Fe), the fitting prioritizes lowering the Ni overproduction. However, it does not falsify models with super-solar Mn/Fe production, because some supernova remnants (SNRs) such as 3C 397 \citep{Mori2018IaYield, Ohshiro2021SNR3C397} exhibit super-solar ratios. The good fit in SNRs and the poor fit in the massive star could imply multiple channels and relative fractions of these SNe. For the same reason, most models overproduce Ni, except for SK18 and LC22, where both are subCh-mass SN Ia models. 

In the top right panel, we show the best-fit models with NKT13 as the massive star models. The Si overproduction, agreeing with previous works, e.g., \cite{Simionescu2019Perseus}, is more prominent. Ar production is relatively lower compared to L25, while the Ca production is, on average, closer to the data. The underproduction of Mn/Fe remains observable below just 1 $\sigma$. The overproduction of Ni in some models is severe, including W7 and PK13. Some Ch-mass (e.g., GC21) and subCh-mass models (SK18 and LC22) remain compatible with the Ni yield.

In the bottom left panel, we use CL04 as the massive star input. The Si production is still more overproduced, similar to NKT13, while S, Ar, and Ca are very close to the values of the Perseus Cluster. There are models where the predicted Mn/Fe can match that in the Perseus Cluster, including W7, PK13, SK18 and GC21, where all except SK18 are Ch-mass models. This supports the nucleosynthesis perspective that the Ch-mass models are necessary components for reconciling the solar value. Most models continue to overproduce Ni by 10-20\% with the exception of the two mentioned subCh-mass models.  

In the bottom right panel, we consider models with WW95 as the massive star input. The overall fitting is similar to the CL04, coincidentally. The fitting of the Si-group elements is well within the 1 $\sigma$ limit. The Mn is underproduced, similar to L25 series, with a similar overproduction in Ni, similar to CL04. 

Despite the fact that we rank the models by the $\chi^2$ value using the chemical composition, models at the two ends are still a very close fit due to the small uncertainties, e.g., that of Ni/Fe is $<10\%$ of the actual value. Meanwhile, theoretical models can easily change by an order of magnitude due to different input physics and progenitor models.

Combining all 4 plots, the higher production of Ar/Fe and Ca/Fe leads to an improved fitting of the Si-group elements compared to \cite{Simionescu2019Perseus}. The Si has less overproduction, while Ar and Ca agree with the data point within the limits. When we examine Mn and Ni, some models could better fit most measured elements, such as SK18. It suggests that the future fitting of the Perseus Cluster will rely on improved SN Ia models. Some SN Ia models, such as SK18, could maintain a high Mn/Fe yield without overproducing Ni. Such a feature is not universally seen in other subCh-mass models such as PK13 and GC21. The GCE models with the lowest $
\chi^2$ value points at a $f_{\rm Chand}$ between $0.2-0.5$. Therefore, it questions whether, despite the wide diversity of SNe Ia observed spectroscopically and photometrically, certain trigger scenarios are more common in the local Universe to fit the Mn and Ni budget.

\subsection{Implied Supernova Rates}
\label{sec:rates}

In Table \ref{tab:bestfit} we listed the SN Ia rates $f_{\rm Ia}$ and the Ch-mass rate $f_{\rm Chand}$ derived from each parameter survey.

In \cite{Timmes1995GCE}, the value of $f_{\rm Ia} \approx 0.007$ is used to create the SN ratio, agreeing with the observed SN Ia rates from nearby galaxies. Our models show a higher value and a wide spread between  0.008 to 0.0435, which translates to the SN Ia rate over SN II rate about $r\sim0.3-0.7$. They agree with some representative observation such as the Palomar Transient Factory \citep[PTF, ][]{Frohmaier2019SNIaRatePTF, Frohmaier2021SNIIRatePTF}, which indicates a value $r\sim0.3-0.5$. This suggests that the Perseus Cluster shares a similar supernova history as nearby galaxies. We remark that the higher $f_{\rm Ia}$ derived here does not necessarily imply a higher \textit{explosion} rate in nearby galaxies. It is because the ICM is the mixture of the ejected gases from the galactic SNe, which requires the gas to escape from the galactic gravitational potential. The contribution of CCSNe, despite their observationally more frequent occurrences than SNe Ia, is suppressed due to their weaker explosions.

The $f_{\rm Chand}$ is much less constrained, where most models populate between $0.2-0.5$, with some models preferring extreme values at pure Ch-mass or subCh-mass models. The mainstream values agree with some recent photometry transient surveys, which report a fraction of $15-41$\%, where an upper limit of 61\% is possible for a more conservative opacity \citep{Bora2024SNIaRate}. 

We note that the more consistent GCE model used in this work also agrees with the estimation in \cite{Simionescu2019Perseus}, which did a direct weighted sum of supernova yields with SN Ia fraction (relative to all SNe) and $f_{\rm Chand}$ being the parameters. In that work, $r=0.2-0.5$ and $f_{\rm Chand}=0.1-0.3$ are reported.

More detailed comparison could be conducted by comparing the elemental trends with stars with various metallicity. This will be useful for future works to explore in details how the supernova models are constrained by multiple observables.

\section{Discussion}
\label{sec:discussion}

\subsection{Future Works}

In this work, we have computed the stellar evolutionary models of $15-60~M_{\odot}$ stars with different metallicity and applied them to the GCE model, using other massive star and SNe Ia explosion models from the literature. 
The new models can overall better reproduce the chemical abundances of the Perseus cluster compared to previous massive star models with some persistent discrepancies noted.
Here we discuss possible directions.


The pair of Si and S poses an interesting question to the massive star models, where the production of these two elements is correlated. The current models suggest that Si is overproduced while S is at its lower side. Further reducing the production of one element will lead to underproduction of the other. This could hint at alternative mechanisms in S production if future models aim at fitting the Si/Fe ratio.



For Ar and Ca, the fit has improved, and all models are now within 1-2 sigma of the Perseus cluster abundances.
However, none of the models can produce above the expectation value. Some other Ca-production channels, e.g., Ca-rich SNe \citep{Polin2021CaRich,Weng2022G306.3,Javier2024CaRichSN}, could be important to fill the missing budget.


Mn and Ni indicate another shortcoming of current models. SNe Ia produce both Mn and Ni, and massive stars produce Ni. Typical SN Ia models report Ni/Fe $\sim 1.5-2.0$. The underproduction of Mn/Fe could have resulted indirectly from the suppression of Ni production in the parameter survey. Future models should reduce the Ni input from both massive stars and SNe Ia such that the Mn/Fe can be more accurately reflected. 


\subsection{Conclusion}

In this article, we investigate how to explain the chemical abundance of the Perseus Cluster. We expanded our calculation in Paper I and computed the massive star evolution and explosive nucleosynthesis for $15 - 60~M_{\odot}$ stars with metallicity $Z=0-1~Z_{\odot}$. We record the thermodynamics and mixing history and post-process it to reconstruct the detailed pre-explosion chemical composition. We use a thermal bomb with a mass cut for the later spherical explosion. We show that the metallicity shows a non-monotonic effect due to the competition of the mass loss, convection history, and the initial composition, where large differences appear in Si-group elements and some in Fe-group elements. Some models, e.g., M20Z1, display large differences from other models due to the shell merger process. 

We apply the massive star models to the GCE code. We update the input yields by our CCSN models from this work and from the literature. We also vary the SN Ia models, including both Ch-mass and subCh-mass models, to identify the effects of these models on the fitting of chemical abundances. We find that whether or not the Perseus Cluster can be well reproduced depends on the good fitting of both channels. For GCE using massive star models from this work, the mass fraction of Si, S, Ar, and Ca against Fe fits better in general, except for Mn/Fe going low by 1-$\sigma$. The overproduction appears in individual elements (S vs Ca), which means that, to further improve the fitting, extra production channels could be necessary. On the other hand, some models predict the models with SN Ia and Ch-mass SN Ia rates the closest to the observations, e.g., NKT13-LN18(Ka4).  

Our work suggests that the matching of the Perseus Cluster requires well-calibrated models of massive stars and SNe Ia. The high-precision measurement, and how the trend of each element could yield unique constraints on the explosion mechanisms of supernovae. In particular, the elements Mn and Ni will provide a strong indication of both SNe Ia and CCSN models. 

\software{  Numpy \citep{Numpy},
            Matplotlib \citep{Matplotlib},
            Pandas \citep{Pandas}
          }

\section*{Acknowledgment}

We thank Frank Timmes for the open-source subroutines of the Helmholtz equation of state, the torch nuclear reaction network, and the Galactic Chemical Evolution code on his \href{https://cococubed.com/}{Cococubed} webpage. We also thank the MESA development team for making the MESA code open-source. 

This material is based upon work supported by the National Science Foundation under Grant AST-2316807.
S.W. and H.Y. thanks for the organization of the SUNY Polytechnic Institute Summer Undergraduate Research Program (SURP) for providing
the opportunity for undergraduate research and development.
K.N. acknowledges support by World Premier International Research Center Initiative (WPI), and JSPS KAKENHI Grant Numbers JP20K04024, JP21H04499, JP23K03452, and JP25K01046. 
A.S. acknowledges the Kavli IPMU for the continued hospitality. SRON Netherlands Institute for Space Research is supported financially by NWO.



\bibliographystyle{aasjournal}
\pagestyle{plain}
\bibliography{biblio}

\end{document}